\documentclass[twocolumn,aps,prb]{revtex4}
\usepackage{graphicx}
\usepackage{amssymb}
\usepackage{amsmath}
\usepackage{bm}
\usepackage{color}

\def\Journal #1,#2,#3,#4#5#6#7{#1 {\bf #2}, #3 (#4#5#6#7)}

\def\e{\varepsilon}
\def\vare{\varepsilon}
\def\p{\prime}
\def\s{\sigma}

\begin{document}

\title{Quantum transport in three-dimensional Weyl electron system}
\author{Yuya Ominato and Mikito Koshino}
\affiliation{Department of Physics, Tohoku University, Sendai 980-8578, Japan}
\date{\today}

\begin{abstract}
Quantum transport in three-dimensional Weyl (massless Dirac) electron system
with long-range Gaussian impurities
is studied theoretically using a self-consistent Born approximation (SCBA).
We find that the conductivity significantly changes its behavior
at a certain critical disorder strength
which separates the weak and strong disorder regimes.
In the weak disorder regime, 
the SCBA conductivity mostly agrees with the Boltzmann conductivity,
except for the Weyl point (the band touching point) at which
the SCBA conductivity exhibits a sharp dip.
In the strong disorder regime, 
the Boltzmann theory fails in all the energy region
and the conductivity becomes larger 
in increasing the disorder potential, contrary to the usual
metallic behavior.
At the Weyl point, the conductivity and the density of states
are exponentially small in the weak disorder regime,
and they abruptly rise at the critical disorder strength.
The qualitative behavior near the zero energy
is well described by an approximate analytic solution of the SCBA
 equation.
The theory applies to three dimensional gapless band structures
including Weyl semimetals.
\end{abstract}

\maketitle

\section{Introduction}
In recent condensed matter physics,
enormous attention has been focused 
on gapless electronic systems where
the conduction band and valence bands touch at some isolated
points in the wave space.
There the electronic band structure is described by
the Weyl equation (the massless Dirac equation),
which leads to unusual physical properties
not observed in conventional metals and semiconductors.
The two-dimensional (2D) version of Weyl electron 
has been extensively investigated 
in graphene \cite{novoselov2005two,zhang2005experimental,ando2005theory,neto2009electronic},
some organic compounds, \cite{katayama2006pressure}
and the surface states in topological insulators. \cite{hasan2010colloquium} 
For three dimensions (3D), there are a number of theoretical proposals
for bulk materials with a gapless band structure
\cite{wan2011topological,young2012dirac,wang2012dirac,singh2012topological,smith2011dirac,liu2013chiral,witczak2012topological,xu2011chern,cho2012possible,halasz2012time}
including Weyl semimetals.

In this paper, we study the electronic transport
in noninteracting 3D Weyl electron in the presence of 
disorder potential.
In the gapless spectrum, generally,
it is a nontrivial task to determine
the conductivity near the Weyl point (band touching point),
where the Boltzmann transport theory fails
and we need to appropriately 
incorporate the finite level broadening effect.
For 2D Weyl electron, 
the transport problem was closely studied,
and the conductivity at the Weyl point was found to be of the order of $e^2/h$
independently of the disorder strength.
\cite{ludwig1994integer,shon1998quantum,ziegler1998delocalization,katsnelson2006zitterbewegung,tworzydlo2006sub,noro2010theory,fradkin1986critical2}
The disorder effect on 3D Weyl electron was studied in
several theoretical works.
\cite{fradkin1986critical2,fradkin1986critical1,burkov2011weyl,burkov2011topological,hosur2012charge,PhysRevLett.112.016402,nandkishore2013dirty,biswas2013diffusive}

Here we calculate the DC conductivity of
3D Weyl electron using a self-consistent Born
approximation (SCBA), which is
one of the theoretical methods to properly treat 
the finite level broadening,
and investigate the dependence of the conductivity
on the Fermi energy and the disorder strength.
In 3D Weyl electron, a short-range disorder 
potential leads to a practical difficulty in which the self-energy 
diverges linearly to the cut-off energy.
To avoid this, we assume long-ranged Gaussian impurities
and achieve the self-consistency, and also study
the dependence of the conductivity
on the characteristic length scale of the impurity potential.

We show that 
the scattering strength is characterized by
a dimensionless parameter $W$
depending on the scattering amplitude and the impurity length scale,
and we find that there is a certain critical disorder strength $W_{\rm c}$
separating the weak and strong disorder regimes.
In the weak disorder regime $(W<W_{\rm c})$, 
the SCBA conductivity mostly agrees with the Boltzmann conductivity
except at the Weyl point, where the conductivity exhibits a sharp dip.
In the strong disorder regime  $(W>W_{\rm c})$, 
the Boltzmann theory fails in all the energy region
and the conductivity becomes larger 
in {\it increasing} the disorder potential, contrary to the usual
metallic behavior.
At the Weyl point, the conductivity and the density of states
are exponentially small in the weak disorder regime,
and they abruptly rise at $W=W_{\rm c}$.
We also show that the qualitative behavior
near the Weyl point
is described by an approximate analytical solution
of the SCBA equation,
where the decay of the impurity matrix element in a large wave number
is approximated by a wave space cut-off.

The paper is organized as follows.
In Sec.\ \ref{sec_form}, we introduce the model Hamiltonian,
and present the formalism to calculate the Boltzmann conductivity
and the SCBA conductivity.
In Sec.\ \ref{sec_app},
we derive an approximate solution of SCBA equation 
near zero energy,
and In Sec.\ \ref{sec_num},
we present the numerical results for the SCBA equation,
and closely argue the behavior of 
the conductivity and the density of states.
A brief summary and discussion are given in Sec.\ \ref{sec_conc}.

\section{Formulation}
\label{sec_form}

\subsection{Hamiltonian}
We consider a three-dimensional, single-node Weyl
electron system
described by a Hamiltonian,
\begin{align}
\mathcal{H}=\hbar v\bm{\s}\cdot\bm{k} + \sum_j U(\bm{r}-\bm{r}_j),
\end{align}
where $\bm{\s}=(\s_x,\s_y,\s_z)$ is the Pauli matrices, $\bm{k}$ is
a wave vector, $v$ is a constant velocity.
The second term is the disorder potential,
where $\bm{r}_j$ is the positions of randomly distributed scatterers.
For each single scatterers, we assume a long-ranged Gaussian potential,
\begin{align}
&U(\bm{r})=\frac{\pm u_0}{(\sqrt{\pi}d_0)^3}
\exp\left(-\frac{r^2}{d_0^2}\right), 
\label{eq_imp_potential}
\end{align}
where $d_0$ is the characteristic length scale,
and scatterers of $\pm u_0$ are randomly distributed with equal probability.
This is Fourier transformed as
$U(\bm{r})=\int{{\rm
d}\bm{q}}u(\bm{q})e^{i\bm{q}\cdot\bm{r}}/{(2\pi)^3}$ where
\begin{align}
&u(\bm{q})=\pm u_0\exp\left(-\frac{q^2}{q_0^2}\right),
\end{align}
and $q_0=2/d_0$.
We introduce an energy scale associated with the potential length scale,
\begin{align}
\e_0=\hbar v q_0.
\end{align}
and define 
a dimensionless parameter characterizing the scattering strength,
\begin{align}
W =\frac{1}{4\pi}\frac{n_{\rm i}u_0^2q_0}{\hbar^2 v^2},
\end{align}
where $n_{\rm i}$ is the number of scatterers per unit volume.



\subsection{Boltzmann transport theory}

The Boltzmann transport equation for the distribution function
$f_{s\bm{k}}$ is given by
\begin{align}
-e\bm{E}\cdot\bm{v}_{s\bm{k}}\frac{\partial f_{s\bm{k}}}{\partial\e_{s\bm{k}}}
        =\sum_{s^\p}\int\frac{{\rm d}\bm{k}^\p}{(2\pi)^3}(f_{s^\p\bm{k}^\p}-f_{s\bm{k}})W_{s^\p\bm{k}^\p,s\bm{k}},
         \label{Boltzmanneq}
\end{align}
where $s=\pm1$ is a label for conduction and valence bands, and
$W_{s^\p\bm{k}^\p,s\bm{k}}$ is the scattering probability,
\begin{align}
W_{s^\p\bm{k}^\p,s\bm{k}}=
&\frac{2\pi}{\hbar}n_{\rm i}
|\langle s' \bm{k}'| U | s \bm{k} \rangle |^2
\delta(\e_{s^\p\bm{k}^\p}-\e_{s\bm{k}}).
\end{align}
The conductivity is obtained by solving Eq.\ (\ref{Boltzmanneq}).
As usual manner, the transport relaxation time 
 $\tau_{\rm tr}$ is defined by
\begin{align}
\frac{1}{\tau_{\rm tr}(\e_{s\bm{k}})}=\int\frac{{\rm d}\bm{k}^\p}{(2\pi)^3}(1-\cos\theta_{\bm{k}\bm{k}^\p})W_{s\bm{k}^\p,s\bm{k}},
\end{align}
where $\theta_{\bm{k}\bm{k}^\p}$ is the angle between $\bm{k}$ and $\bm{k}^\p$.
For the isotropic scatterers, i.e., $u(\bm{q})$ 
depending only on $q=|\bm{{q}}|$, 
it is straightforward to show that $\tau_{\rm tr}(\e_{s\bm{k}})$ solely depends
on the energy $\e$ and written as \cite{burkov2011topological}
\begin{align}
\frac{1}{\tau_{\rm tr}(\e)}=&
\frac{\pi}{\hbar}n_{\rm i} D_0(\e)
\int_{-1}^{1} d(\cos\theta) \, 
u^2[2k\sin(\theta/2)]
\notag\\
&\qquad\qquad \times
(1-\cos\theta)\frac{1+\cos\theta}{2},
\end{align}
where $k = \e/(\hbar v)$ and
$D_0(\e)$ is the density of states in the ideal Weyl electron,
\begin{align}
D_0(\e)=\frac{\e^2}{2\pi^2(\hbar v)^3}.
\end{align}
The conductivity at $T=0$ is written as
\begin{align}
\s_{\rm B}(\e)=e^2\frac{v^2}{3}D_0(\e)\tau_{\rm tr}(\e),
\end{align}

\begin{figure}
\begin{center}
\leavevmode\includegraphics[width=1.0\hsize]{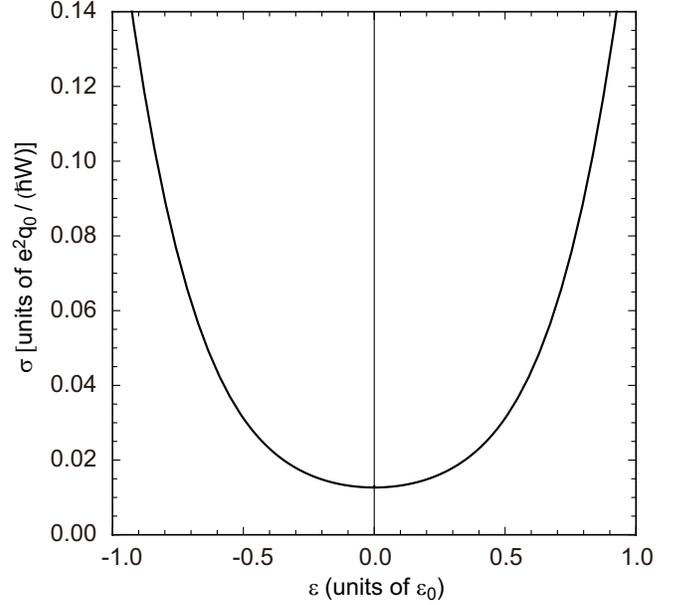}
\end{center}
\caption{Boltzmann conductivity [Eq.\ (\ref{Boltzmann})]
plotted as a function of the Fermi energy.
}
\label{boltz}
\end{figure}

For the Gaussian scatter,
Eq.\ (\ref{eq_imp_potential}), 
the relaxation time and conductivity are explicitly written as
\begin{align}
&\tau_{\rm tr}(\e)=\frac{\hbar}{2 \e_0 W}
\,h\left(\frac{\e}{\e_0}\right), \notag \\
&\s_{\rm B}(\e)=\frac{1}{12\pi^2}\frac{e^2
 q_0}{\hbar}\frac{1}{W}\left(\frac{\e}{\e_0}\right)^2
h\left(\frac{\e}{\e_0}\right), \label{Boltzmann}
\end{align}
where
\begin{align}
&h(x)=\frac{64x^4}{4x^2-1+(4x^2+1)\exp(-8x^2)}.
\end{align}
In particular, the conductivity at the Weyl point is
\begin{align}
\s_{\rm B}(0)=\frac{1}{8\pi^2}\frac{e^2 q_0}{\hbar}\frac{1}{W}
      =\frac{1}{2\pi}\frac{e^2 v^2\hbar}{n_{\rm i}u_0^2},
\label{eq_boltz_min}
\end{align}
which is independent of $q_0$.
\cite{burkov2011topological,hosur2012charge}
Fig.\ \ref{boltz} shows the Boltzmann conductivity Eq.\ (\ref{Boltzmann})
as a function of the Fermi energy.

\subsection{Self-consistent Born approximation}

We introduce the self-consistent Born approximation (SCBA)
for 3D Weyl electron 
in a similar manner to the 2D version in Ref. \cite{noro2010theory}.
The following formulation does not depend on 
the specific form of the single impurity potential $U(\bm{r})$,
as long as it is isotropic.
We define the averaged Green's function as
\begin{align}
\hat G(\bm{k},\e) = \biggl\langle\frac{1}{\e-\mathcal{H}}\biggr\rangle 
       = \frac{1}{\e - \hbar v\bm{\sigma}\cdot\bm{k} - \hat \Sigma(\bm{k},\e)},
\label{eq_green}
\end{align}
where $\langle\cdots\rangle$ represents the average over
the configuration of the impurity position.
$\hat\Sigma(\bm{k},\e)$ is the self-energy matrix,
which is approximated in SCBA as
\begin{align}
\hat \Sigma(\bm{k},\e)=\int\frac{{\rm d}\bm{k}^\p}{(2\pi)^3}n_{\rm
 i}|u(\bm{k}-\bm{k}^\p)|^2\hat G(\bm{k}^\p,\e).
\label{eq_self}
\end{align}
Eqs.\ (\ref{eq_green}) and (\ref{eq_self})
are a set of equations to be solved self-consistently.
From the symmetry of the present system, the self-energy matrix can
be expressed as

\begin{align}
\hat \Sigma(\bm{k},\e)=\Sigma_1(k,\e)+\Sigma_2(k,\e)(\bm{\sigma}\cdot\bm{n}),
\end{align}
where $k=|\bm{k}|$ and $\bm{n}=\bm{k}/k$.
By defining $X(k,\e)$ and $Y(k,\e)$ as
\begin{align}
X(k,\e)&=\e-\Sigma_1(k,\e), \\ 
Y(k,\e)&=\hbar v k +\Sigma_2(k,\e),
\end{align}
Eqs.\ (\ref{eq_green}) and (\ref{eq_self}) are written as
\begin{align}
\hat G(\bm{k},\e)= \frac{1}{X(k,\e)-Y(k,\e)(\bm{\sigma}\cdot\bm{n})},
\end{align}
and 
\begin{align}
\hat \Sigma(\bm{k},\e)=\int\frac{{\rm d}\bm{k}^\p}{(2\pi)^3}n_{\rm i}|u(\bm{k}-\bm{k}^\p)|^2
                           \frac{X^\p+Y^\p(\bm{\sigma}\cdot\bm{n}^\p)}{{{X^\p}^2-{Y^\p}^2}} \label{self1}
\end{align}
where $X^\p=X(k^\p,\e)$, $Y^\p=Y(k^\p,\e)$, and $\bm{n}^\p=\bm{k}^\p/k^\p$.

Now, we divide $\bm{n}^\p$ as
\begin{align}
\bm{n}^\p=\bm{n}^\p_\parallel+\bm{n}^\p_\perp.
\end{align}
where $\bm{n}^\p_\parallel=(\bm{n}\cdot\bm{n}^\p)\bm{n}$ 
is the component of parallel to $\bm{n}$,
and $\bm{n}^\p_\perp$ is the perpendicular part.
Then Eq.\ (\ref{self1}) becomes
\begin{align}
\hat \Sigma(\bm{k},\e)=&\int\frac{{\rm d}\bm{k}^\p}{(2\pi)^3}n_{\rm i}|u(\bm{k}-\bm{k}^\p)|^2
                                      \frac{X^\p}{{{X^\p}^2-{Y^\p}^2}} \notag \\
                                 +&\int\frac{{\rm d}\bm{k}^\p}{(2\pi)^3}n_{\rm i}|u(\bm{k}-\bm{k}^\p)|^2
                                      \frac{Y^\p}{{{X^\p}^2-{Y^\p}^2}}(\bm{\s}\cdot\bm{n}^\p_\parallel) \notag \\
                                 +&\int\frac{{\rm d}\bm{k}^\p}{(2\pi)^3}n_{\rm i}|u(\bm{k}-\bm{k}^\p)|^2
                                      \frac{Y^\p}{{{X^\p}^2-{Y^\p}^2}}(\bm{\s}\cdot\bm{n}^\p_\perp).
\end{align}
The third term vanishes after the integration over the $\bm{k}^\p$ direction,
giving
\begin{align}
\hat \Sigma(\bm{k},\e)=&\int_0^\infty\frac{{k^\p}^2{\rm d}k^\p}{(2\pi)^3}n_{\rm i}V_0^2(k,k^\p)
                                      \frac{X^\p}{{{X^\p}^2-{Y^\p}^2}} \notag \\
                                 +&(\bm{\sigma}\cdot\bm{n})
                                     \int_0^\infty\frac{{k^\p}^2{\rm d}k^\p}{(2\pi)^3}n_{\rm i}V_1^2(k,k^\p)
                                      \frac{Y^\p}{{{X^\p}^2-{Y^\p}^2}},
\label{self2}
\end{align}
where 
\begin{align}
V_n^2(k,k^\p)=2\pi\int^1_{-1}{\rm d}(\cos\theta_{\bm{k}\bm{k}^\p})
                     |u(\bm{k}-\bm{k}^\p)|^2\cos^n\theta_{\bm{k}\bm{k}^\p}. \label{vn}
\end{align}
Eq.\ (\ref{self2}) immediately leads to
the self-consistent equation,
\begin{align}
X(k,\e)&=\e-\int_0^\infty\frac{{k^\p}^2{\rm d}k^\p}{(2\pi)^3}n_{\rm i}V_0^2(k,k^\p)
                                      \frac{X^\p}{{{X^\p}^2-{Y^\p}^2}}, 
\notag \\
Y(k,\e)&=\hbar v k +\int_0^\infty\frac{{k^\p}^2{\rm d}k^\p}{(2\pi)^3}n_{\rm i}V_1^2(k,k^\p)
                                      \frac{Y^\p}{{{X^\p}^2-{Y^\p}^2}},
\label{self_xy}
\end{align}
which are to be solved numerically.
From the obtained Green's function,
the density of states per unit area is calculated as
\begin{align}
D(\e)=-\frac{1}{\pi}{\rm Im}\int\frac{{\rm d}\bm{k}}{(2\pi)^3}{\rm Tr}[\hat G(\bm{k},\e+i0)].
\end{align}

The Kubo formula for the conductivity is given by
\begin{align}
\sigma(\e)&=-\frac{\hbar e^2v^2}{4\pi}\sum_{s,s^\p=\pm1}s
 s^\p\int\frac{{\rm d}\bm{k}^\p}{(2\pi)^3}  
{\rm Tr}\biggl[
\s_x\hat G(\bm{k}^\p,\e+is0) \notag \\
&{~~~}\times\hat J_x(\bm{k}^\p,\e+is0,\e+is^\p0)\hat
 G(\bm{k}^\p,\e+is^\p0)
\biggr],
\end{align}
where $\hat J_x$ is current vertex-part satisfying the Bethe-Salpeter equation
\begin{align}
\hat J_x(\bm{k},\e,\e^\p)=\sigma_x+&\int\frac{{\rm d}\bm{k}^\p}{(2\pi)^3}
                                                   n_{\rm i}|u(\bm{k}-\bm{k}^\p)|^2\hat G(\bm{k}^\p,\e) \notag \\
&\times\hat J_x(\bm{k}^\p,\e,\e^\p)\hat G(\bm{k}^\p,\e^\p). \label{GJG}
\end{align}
To calculate this, we consider an integral
\begin{align}
I(\bm{k})=\int\frac{{\rm d}\bm{k}^\p}{(2\pi)^3}|u(\bm{k}-\bm{k}^\p)|^2F(k^\p)
                   (\bm{\s}\cdot\bm{n}^\p)\s_x(\bm{\s}\cdot\bm{n}^\p),
\end{align}
where $F(k)$ is an arbitrary function.
After some algebra, we have
\begin{align}
I(\bm{k})=&\s_x\int\frac{{k^\p}^2{\rm d}k^\p}{(2\pi)^3}F(k^\p) 
\left(-\frac{1}{2}V_0^2(k,k^\p)+\frac{1}{2}V_2^2(k,k^\p)\right) \notag \\
             &+(\bm{\s}\cdot\bm{n})\s_x(\bm{\s}\cdot\bm{n})\int\frac{{k^\p}^2{\rm d}k^\p}{(2\pi)^3}F(k^\p) \notag \\
& \qquad\qquad
\times\left(-\frac{1}{2}V_0^2(k,k^\p)+\frac{3}{2}V_2^2(k,k^\p)\right).
\end{align}
In a similar way as for the self-energy, we obtain
\begin{align}
 &\int\frac{{\rm
 d}\bm{k}^\p}{(2\pi)^3}|u(\bm{k}-\bm{k}^\p)|^2F(k^\p)(\bm{\s}\cdot\bm{n}^\p)\s_x
 \notag \\
& \qquad =(\bm{\s}\cdot\bm{n})\s_x\int\frac{{k^\p}^2{\rm d}k^\p}{(2\pi)^3}F(k^\p)V_1^2(k,k^\p), \notag \\
 &\int\frac{{\rm d}\bm{k}^\p}{(2\pi)^3}|u(\bm{k}-\bm{k}^\p)|^2F(k^\p)\s_x(\bm{\s}\cdot\bm{n}^\p) \notag \\
& \qquad = \s_x(\bm{\s}\cdot\bm{n})\int\frac{{k^\p}^2{\rm d}k^\p}{(2\pi)^3}F(k^\p)V_1^2(k,k^\p).
\end{align}
Using these, the vertex part $\hat J$ is written as
\begin{align}
\hat J_x(\bm{k},\e,\e^\p)=\s_x J_0(k,\e,\e^\p)+(\bm{\s}\cdot\bm{n})\s_x(\bm{\s}\cdot\bm{n}) J_1(k,\e,\e^\p) \notag \\
                           +(\bm{\s}\cdot\bm{n})\s_x J_2(k,\e,\e^\p)+\s_x(\bm{\s}\cdot\bm{n})J_3(k,\e,\e^\p), \label{J}
\end{align}
and the Bethe-Salpeter equation becomes
\begin{align}
   \begin{pmatrix}
    J_0  \\
    J_1  \\
    J_2  \\
    J_3   
   \end{pmatrix}=\begin{pmatrix}
                        1  \\
                        0  \\
                        0  \\
                        0   
                       \end{pmatrix}+&\int_0^\infty\frac{{k^\p}^2{\rm d}k^\p}{(2\pi)^3}
                                           \frac{n_{\rm i}}{(X^2-Y^2)({X^\p}^2-{Y^\p}^2)} \notag \\
            &\times\begin{pmatrix}
                        V_0^2 & -(V_0^2-V_2^2)/2 & 0 & 0 \\
                        0 & -(V_0^2-3V_2^2)/2  & 0 & 0 \\
                        0 & 0 & V_1^2 & 0 \\
                        0 & 0 & 0 & V_1^2
                       \end{pmatrix} \notag \\
            &\times\begin{pmatrix}
                        XX^\p & YY^\p & YX^\p & XY^\p \\
                        YY^\p & XX^\p & XY^\p & YX^\p \\
                        YX^\p & XY^\p & XX^\p & YY^\p \\
                        XY^\p & YX^\p & YY^\p & XX^\p
                       \end{pmatrix}    \begin{pmatrix}
                                              J_0^\p  \\
                                              J_1^\p  \\
                                              J_2^\p  \\
                                              J_3^\p       
                                             \end{pmatrix},
\label{self_J}
\end{align}
where $X=X(k^\p,\e)$, $X^\p=X(k^\p,\e^\p)$, $J_0=J_0(k,\e,\e^\p)$, $J_0^\p=J_0(k^\p,\e,\e^\p)$, etc.
Finally, the conductivity is written as 
\begin{align}
\s(\e)=&\frac{4\hbar e^2v^2}{3}\int_0^\infty\frac{k^2{\rm d}k}{(2\pi)^3} \notag \\
                     \times{\rm Re}&\biggl[\frac{1}{|X^2-Y^2|^2} \notag \\
                             &\times\Bigl\{(3|X|^2-|Y|^2)J_0^{+-}+(3|Y|^2-|X|^2)J_1^{+-} \notag \\
                             &+(3YX^\ast-XY^\ast)J_2^{+-}+(3XY^\ast-YX^\ast)J_3^{+-}\Bigr\} \notag \\
                             &-\frac{1}{(X^2-Y^2)^2} \notag \\
                             &\times\Bigl\{(3X^2-Y^2)J_0^{++}+(3Y^2-X^2)J_1^{++} \notag \\
                             &+2XYJ_2^{++}+2XYJ_3^{++}\Bigr\}\biggr], \label{cond}
\end{align}
where $X=X(k,\e+i0)$, $J_0^{ss'} = J_0(k,\e+is0,\e+is'0)$, etc. 


The SCBA is a valid approximation when
the disorder scattering is relatively weak so that
$k_F l \gg 1$,  where $k_F$ is the Fermi wave length, and 
$l = v_F \tau$ is the mean free path given by
the Fermi velocity $v_F$ and the relaxation time $\tau$.
In the 3D Weyl electron, $v_F$ is the constant band velocity $v$,
and in the case of Gaussian impurities,
$\tau$ is roughly estimated by
$\tau_{\rm tr}$ in the Boltzmann theory, Eq.\ (\ref{Boltzmann}).
The condition then becomes
\begin{equation}
 \frac{1}{2}
\left(
\frac{\e}{\e_0}\right)
h\left(
\frac{\e}{\e_0}\right)
\gg W.
\label{eq_scba_condition}
\end{equation}
Near zero energy $\e \ll \e_0$, in particular,
$h(\e/\e_0)$ is approximated by $(3/2)(\e/\e_0)^{-2}$ 
and the condition reduces to $(4/3)\e/\e_0 \ll 1/W$, i.e.,
the approximation is better for smaller energy.
In higher energy region $\e > \e_0$,
the function $h(\e/\e_0)$ approximates $16(\e/\e_0)^2$,
and the condition becomes $8(\e/\e_0)^3 \gg W$,
i.e., the approximation is valid also in the higher energy region.

\section{Approximate analytical solution near zero energy}
\label{sec_app}

Near $\vare = 0$,
we can derive an approximate analytical solution
of the SCBA equation in the previous section,
as long as the level broadening $\Gamma(\e)$
is much smaller than $\vare_0$.
There we replace $u(q)$ with the constant $u_0$
(i.e., short-ranged impurity),
but, instead, introduce a cutoff $k_{\rm c} \sim q_0$ 
in the $k$-space integral 
to simulate the exponential decay of $u(q)$ in a large $q$.
The approximation is rather crude,
while it effectively explains the qualitative behavior 
peculiar to the 3D Weyl electron as shown in the following section.

In this simplified system (short-range impurities with cutoff),
the self-energy equation Eq.\ (\ref{eq_self}) is approximately solved 
in $\e \ll \e_{\rm c}$ and $\Gamma \ll \e_{\rm c}$
as
\begin{align}
\hat \Sigma(\bm{k},\e) \approx (1-\alpha)\e - i\Gamma(\e),
\label{xy_approx}
\end{align}
where
\begin{align}
\Gamma(\e)
&=\frac{\Gamma_W}{2}
    +\sqrt{
             \left(
                  \frac{\Gamma_W}{2}
             \right)^2
             +\alpha^2\e^2
      },
\notag \\
\alpha
&=\left(
          1-\frac{W}{W_{\rm c}}
    \right)^{-1}, 
\notag \\
\Gamma_W
&=\e_0
    \left(
         \frac{1}{W_{\rm c}}-\frac{1}{W} 
    \right),
\label{gamma_approx}
\end{align}
and 
\begin{align}
W_{\rm c}
&=\frac{\pi}{2}
    \frac{\e_0}{\e_{\rm c}},
\label{gamma_approx2}
\end{align}
where $\vare_c = \hbar v k_c$ is the cut-off energy.
The density of states is then written in terms of $\Gamma(\e)$ as
\begin{align}
D(\e)=\frac{\e_0}{2\pi^2(\hbar v)^3}\frac{\Gamma(\e)}{W}.
\label{dos_app}
\end{align}

At $\vare=0$, in particular, $\Gamma(\vare)$ becomes
\begin{align}
\Gamma(0) =\begin{cases}
                0 & (W<W_{\rm c}) \\                      
               \Gamma_W & (W>W_{\rm c})
              \end{cases},
\label{eq_gamma_approx}
\end{align}
i.e., the self-energy, and thus the density of states,
become zero in the weak disorder regime $W<W_{\rm c}$,
and abruptly rise in the strong disorder regime $W>W_{\rm c}$.
The vanishing $\Gamma(0)$ at a finite $W$ is peculiar to three dimensions, 
and it is intuitively understood as follows.
By assuming a solution of the form Eq.\ (\ref{xy_approx}),
the first equation in Eq.\ (\ref{self_xy}) 
can be written at $\vare = 0$ as
\begin{align}
\Gamma = 
&\frac{n_{\rm i}u_0^2}{2\pi^2}
\int^{k_{\rm c}}_0k^2{\rm d}k \frac{\Gamma}{(\hbar v k)^2 + \Gamma^2},
\label{eq_self_app_e0}
\end{align}
which is to be solved for $\Gamma$.
For a non-zero $\Gamma$, it becomes
\begin{align}
1 = 
&\frac{n_{\rm i}u_0^2}{2\pi^2}
\int^{k_{\rm c}}_0k^2{\rm d}k \frac{1}{(\hbar v k)^2 + \Gamma^2}.
\label{eq_self_app_e0_2}
\end{align}
When the right-hand side of Eq.\ (\ref{eq_self_app_e0_2})
is viewed as a function of $\Gamma$,
it has an upper bound $n_{\rm i}u_0^2k_c/(2\pi^2\hbar^2v^{2})$,
which is achieved at $\Gamma = 0$. 
When the scattering strength
is so small that $n_{\rm i}u_0^2k_c/(2\pi^2\hbar^2v^{2})$ is smaller
than 1, Eq.\ (\ref{eq_self_app_e0_2}) has no solution,
and we are left only with a trivial solution $\Gamma=0$ 
in Eq.\ (\ref{eq_self_app_e0}).
This critical condition exactly corresponds to $W < W_{\rm c}$.
In contrast, the self-consistent equation in 2D
always has a non-zero solution 
for any scattering strength, \cite{shon1998quantum}
because in Eq.\ (\ref{eq_self_app_e0_2}), 
$k^2 dk$ is replaced with $k dk$,
and then the right-hand side 
logarithmically diverges in $\Gamma \to 0$,
giving no upper bound.


In a similar manner, the Bethe-Salpeter equation Eq.\ (\ref{self_J}) 
is approximately solved as
\begin{align}
   \begin{pmatrix}
    J_0  \\
    J_1  \\
    J_2  \\
    J_3   
   \end{pmatrix}& \approx \begin{pmatrix}
                        J  \\
                        0  \\
                        0  \\
                        0   
                       \end{pmatrix},
\label{vertex_approx}
\end{align}
where
\begin{align}
J&=\left[1+\frac{1}{3}\frac{W}{W_{\rm c}}\right]^{-1}.
\label{J_approx}
\end{align}
The conductivity is obtained as
\begin{align}
\sigma(\e)=\frac{J}{12\pi^2}\frac{e^2}{\hbar}
\frac{1}{\hbar v}\frac{3\Gamma(\e)^2+\alpha^2\e^2}{\Gamma(\e)}.
\label{cond_app}
\end{align}

When $\Gamma(0)$ is non-zero,  
the conductivity at $\e=0$ can be simply obtained by replacing 
$\Gamma(\e)$ with $\Gamma(0)$ as
\begin{align}
\sigma(0)=
\frac{J}{4\pi^2}\frac{e^2}{\hbar}
\frac{\Gamma(0)}{\hbar v}.
\label{cond_app_e0}
\end{align}
In the strong disorder regime $W>W_{\rm c}$,
this gives
\begin{align}
\sigma(0)=
 \frac{J}{12\pi^2}\frac{e^2q_0}{\hbar}\times
           3\left(\dfrac{1}{W_{\rm c}}-\dfrac{1}{W}\right) \quad (W>W_{\rm c}).
\label{cond_approx_zero_1}
\end{align}
In the weak disorder regime $W<W_{\rm c}$, 
Eq.\ (\ref{cond_app_e0}) is no longer valid since $\Gamma(0)=0$, 
and then we need to take a limit $\e\to 0$ to consider $\sigma(0)$.
We expand $\Gamma(\e)$ in Eq.\ (\ref{gamma_approx}) as
\begin{align}
\Gamma(\e) \approx 
                \dfrac{\alpha^2\e^2}{|\Gamma_W|} \quad (W<W_{\rm c}),
\label{eq_gamma_approx_near_zero}
\end{align}
and obtain
\begin{align}
\lim_{\e\to 0}\sigma(\e)=
 \frac{J}{12\pi^2}\frac{e^2q_0}{\hbar}\times
           \left(\dfrac{1}{W}-\dfrac{1}{W_{\rm c}}\right) \quad (W<W_{\rm c}).
\label{cond_approx_zero_2}
\end{align}
From Eqs.\ (\ref{cond_approx_zero_1}) and (\ref{cond_approx_zero_2}),
we see that the Weyl-point conductivity $\sigma(0)$ vanishes at
$W=W_{\rm c}$, and it increases as 
$W$ goes away from $W_{\rm c}$ in either direction.

{
In the vicinity of the critical point $W=W_{\rm c}$,
$\alpha$ becomes large and $\Gamma(\e)$ and $\sigma(\e)$
approximate
linear functions,
\begin{eqnarray}
\begin{array}{l}
\Gamma(\e) \approx |\alpha \e|
\\
\displaystyle \sigma(\e) \approx \frac{1}{4\pi^2}\frac{e^2}{\hbar} \frac{|\alpha \e|}{\hbar v} 
\end{array}
\quad (W \approx W_{\rm c}),
\label{eq_gamma_at_crit}
\end{eqnarray}
of which gradient diverges at $W=W_{\rm c}$.
}

\begin{figure}
\begin{center}
\leavevmode\includegraphics[width=0.9\hsize]{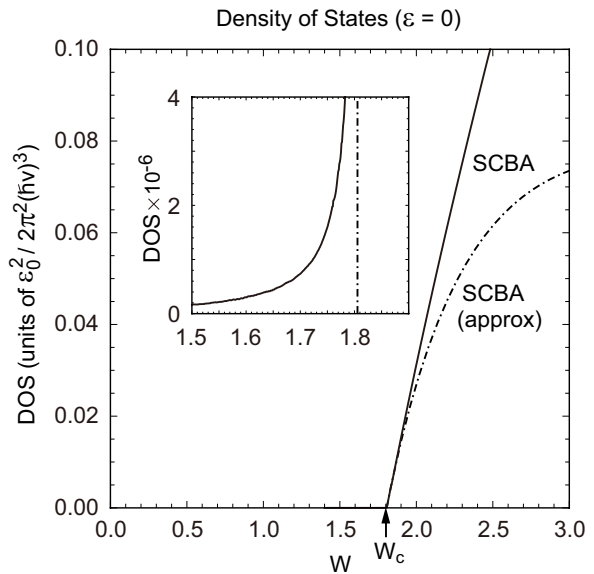}
\end{center}
\caption{(Solid) Density of states at zero energy
calculated by the SCBA.
(Dotted-dashed) Approximate expression Eq.\ (\ref{dos_app})
with Eq.\ (\ref{gamma_approx}),
where $\vare_c/\vare_0$ is taken as $0.87$.
Inset shows the density of states
near $W_{\rm c}\approx1.806$ in a smaller scale.
}
\label{dos}
\end{figure}

\begin{figure*}
\begin{center}
\leavevmode\includegraphics[width=0.9\hsize]{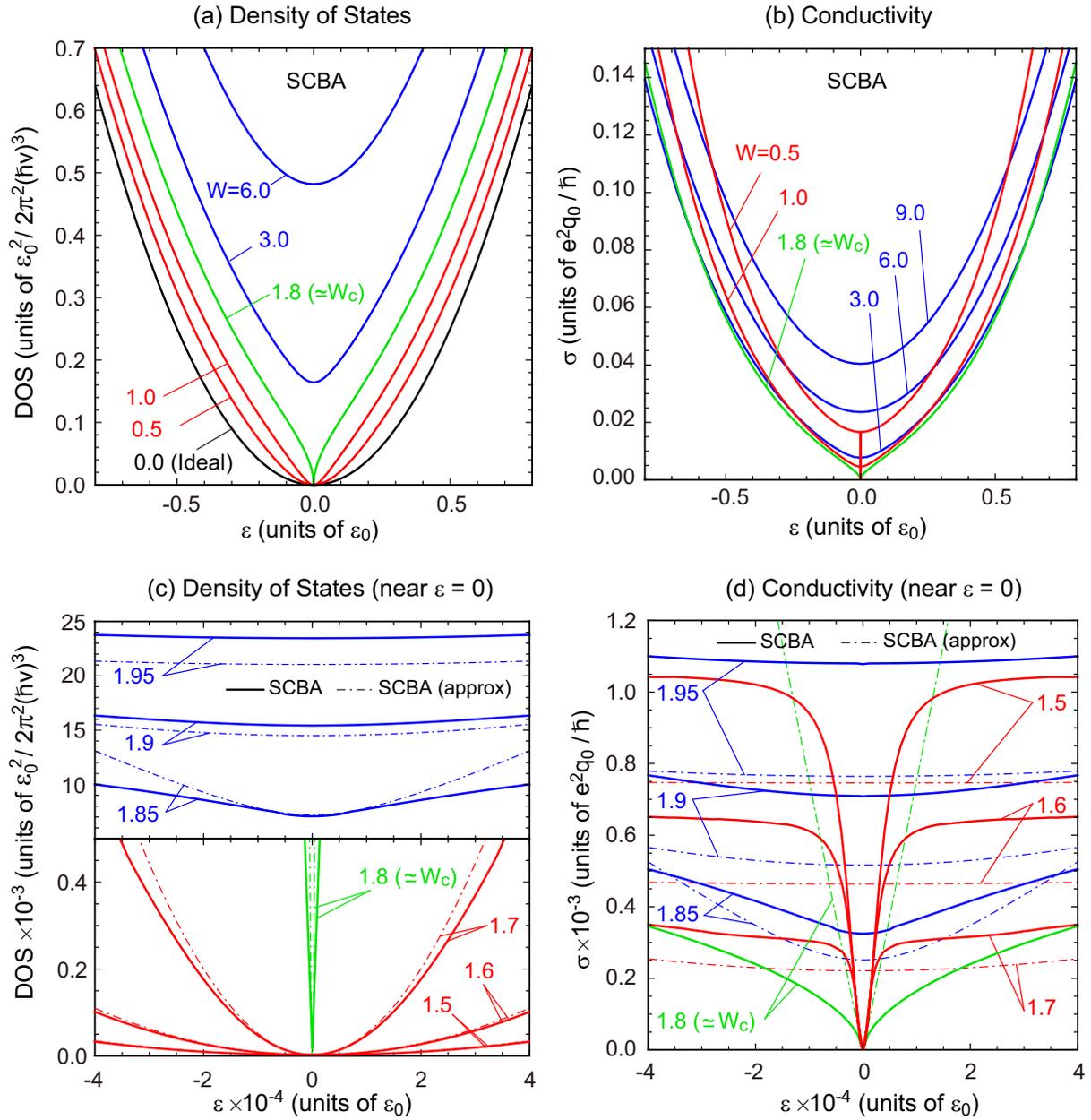}
\end{center}
\caption{Density of states (a,c) and the conductivity (b,d) 
calculated by the SCBA, as a function of the Fermi energy.
The panels (c) and (d) show the detailed plots near $\e=0$
of (a) and (b), respectively.
The dotted-dashed line in Fig.\ \ref{fe}(c) and (d)
represent the approximate SCBA solution,
Eq.\ (\ref{dos_app}) and Eq.\ (\ref{cond_app}), respectively.
}
\label{fe}
\end{figure*}

\begin{figure}
\begin{center}
\leavevmode\includegraphics[width=0.9\hsize]{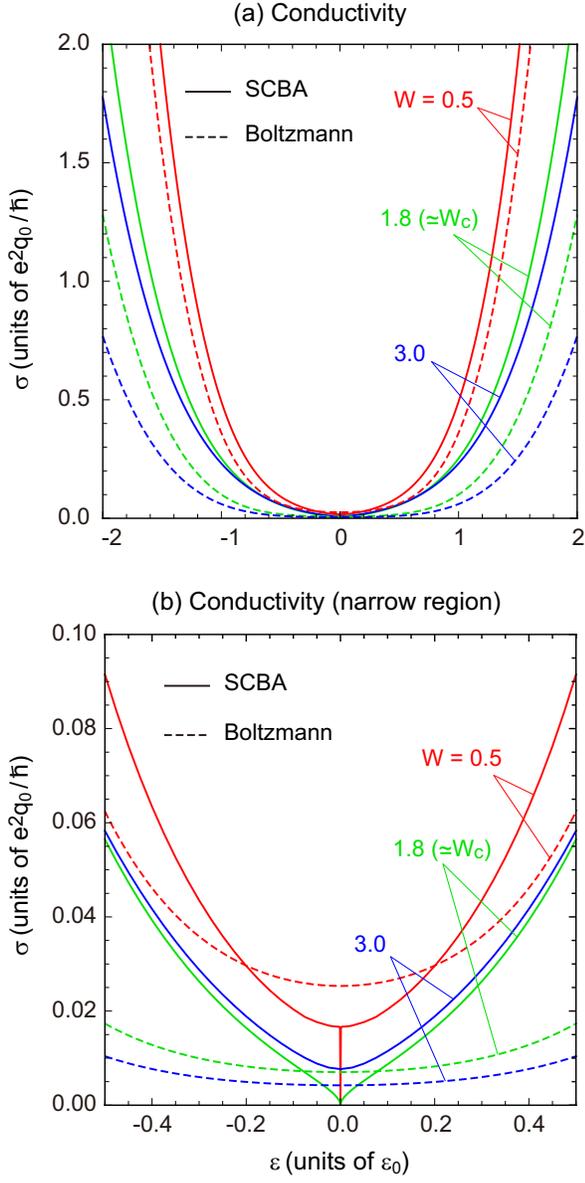}
\end{center}
\caption{
SCBA conductivity and the Boltzmann conductivity 
[Eq.\ (\ref{Boltzmann})] 
at several $W$'s plotted for (a) wide and (b) narrow energy regions.
}
\label{comp}
\end{figure}

\begin{figure}
\begin{center}
\leavevmode\includegraphics[width=0.9\hsize]{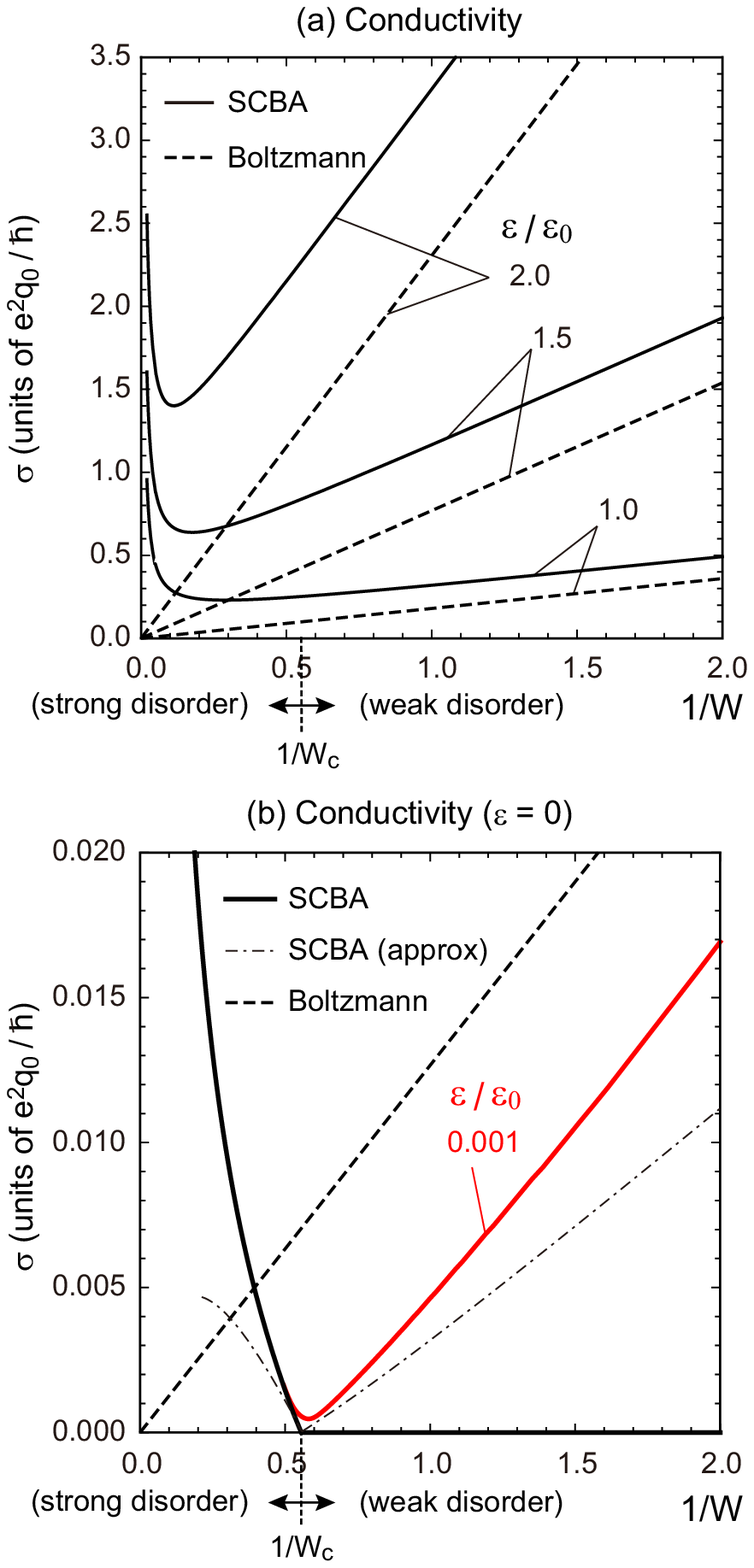}
\end{center}
\caption{
(a) SCBA conductivity (solid) and the Boltzmann conductivity (dashed) 
as a function of $1/W$, for some fixed Fermi energies. 
(b) A similar plot 
for small Fermi energies, $\e=0$ and $0.001\e_0$.
Dotted-dashed line represents
the approximate SCBA solution, 
Eqs.\ (\ref{cond_approx_zero_1}) and (\ref{cond_approx_zero_2})
at $\e=0$.
}
\label{imp}
\end{figure}

\section{Numerical Results}
\label{sec_num}


We solve the SCBA equations Eq.\ (\ref{self_xy}) and (\ref{self_J})
by numerical iteration and calculate the density of states 
and the conductivity.
Fig.\ \ref{dos} shows the density of states  at $\e=0$
as a function of $W$.
The behavior is qualitatively similar to
the approximate analysis in the previous section.
There is a critical disorder strength $W_{\rm c} \approx 1.806$,
and the density of states rapidly increases once
$W$ enters the strong disorder regime $W>W_{\rm c}$.
The dotted-dashed line in Fig.\ \ref{dos} shows
the approximate SCBA solution of Eq.\ (\ref{dos_app}) at $\e=0$,
where $\e_{\rm c}/\e_0$ is taken as $\approx 0.87$
to fit $W_{\rm c}$ to the numerically obtained value.
It nicely reproduces the increase in $W>W_{\rm c}$,
though the approximation fails in larger $W$ because the assumption
$\Gamma\ll\e_{\rm c}$ in deriving Eq.\ (\ref{gamma_approx})
becomes no longer valid as $\Gamma$ increases.
Actually, the density of states in the weak disorder regime $W<W_{\rm c}$
does not completely vanish in the numerics unlike the analytic approximation,
but an exponentially small value
remains as shown in the inset in Fig.\ \ref{dos}.
The rapid increase in $W<W_{\rm c}$ is roughly expressed by
$\propto 1/(W_{\rm c}-W)$.
As we will argue later, this small residue 
leads to a significant difference 
in the zero-energy conductivity between
the numerical calculation and the analytic approximation.


Figs.\ \ref{fe} (a) and (b) show 
the density of states and the conductivity 
as a function of the Fermi energy, respectively,
which are numerically calculated by the SCBA.
Figs.\ \ref{fe} (c) and (d) are the detailed plots around zero energy
for Figs.\ \ref{fe} (a) and (b), respectively.
We see that the density of states 
is enhanced in all the energy region 
with the increase of the scattering strength $W$.
In the weak disorder regime $W<W_{\rm c}$,
it approximates a quadratic curve in the vicinity of $\e=0$,
and it nearly sticks to zero at the origin.
At the critical point $W_{\rm c}$, the curve exhibits a wedge-like shape,
and in the strong disorder regime $W>W_{\rm c}$, 
the bottom of the curve departs from zero as already argued.
The dotted-dashed line in Fig.\ \ref{fe}(c) represents
the approximate SCBA solution near zero energy, 
Eq.\ (\ref{dos_app}). 
It reproduces the qualitative behavior
of the numerical curve.
{
At $W=1.8$, slightly away from
the critical point $W_{\rm c}\approx 1.806$,
the density of states is approximated by
a linear function in accordance with Eq.\ (\ref{eq_gamma_at_crit}).
}


The conductivity exhibits significantly different behaviors
between the weak and strong disorder regimes.
In Fig.\ \ref{comp}, we compare the SCBA conductivity
and the Boltzmann conductivity Eq.\ (\ref{Boltzmann}) at several $W$'s
in (a) wide and (b) narrow energy regions.
We see that
the SCBA agrees well with the Boltzmann conductivity in small $W$,
while the discrepancy becomes significant as $W$ increases.
The Boltzmann theory works well 
when the condition that the self-energy is much smaller than
the Fermi energy, so that
the theory naturally stands in the weak disorder regime.
In the strong disorder regime $W>W_{\rm c}$,
the Boltzmann approximation fails
and the SCBA conductivity is {\it enhanced} in increasing $W$
as observed in Figs.\ \ref{fe} (b) and (d),
contrary to the usual metallic behavior.
When $W$ is too strong to violate the condition
Eq.\ (\ref{eq_scba_condition}), the SCBA is no longer valid
and the correction from the quantum interference effect 
would be important.

In the weak disorder regime $W<W_{\rm c}$,
we notice that the conductivity $\sigma(\vare)$
exhibits a sharp dip at the zero energy,
as shown in Fig.\ \ref{fe}(d) in a greater scale.
The dotted-dashed line in Fig.\ \ref{fe}(d) indicates
the approximate SCBA solution near zero energy, 
Eq.\ (\ref{cond_app}),
with $\e_{\rm c}/\e_0 = 0.87$ (the same value used in the DOS plot).
We see that the analytic approximation
fails to describe the zero-energy dip observed in the numerics,
while, in the flat region outside the dip, 
the approximation qualitatively reproduces the $W$-dependence
of the numerical conductivity.

The conductivity dip actually originates from 
an exponentially small self-energy remaining at $\e=0$, 
which is missing in the analytic approximation.
Indeed, the entire curve including the dip
is qualitatively reproduced by Eq.\ (\ref{cond_app}),
when $\Gamma(\e)$ of Eq.\ (\ref{eq_gamma_approx_near_zero}) 
is modified by
\begin{align}
\Gamma(\e) \approx \Gamma_{\rm num}+
\dfrac{\alpha^2\e^2}{|\Gamma_W|} \quad (W<W_{\rm c}),
\label{eq_gamma_approx_near_zero_corr}
\end{align}
where $\Gamma_{\rm num}$
is the small residue of the selfenergy  at $\e=0$
in the numerical calculation.
The conductivity $\sigma(0)$ is then given by Eq.\ (\ref{cond_app_e0}),
and thus is exponentially small.
In increasing the Fermi energy $\e$,
$\Gamma_{\rm num}$ becomes less important 
in Eq.\ (\ref{eq_gamma_approx_near_zero_corr}),
and the conductivity gradually approaches the original analytic
expression Eq.\ (\ref{cond_approx_zero_2}).
The energy width of the dip is roughly estimated by the condition
$\alpha^2\e^2/|\Gamma_W| \sim \Gamma_{\rm num}$.


In Fig.\ \ref{imp}(a), 
the SCBA conductivity (solid) and the Boltzmann conductivity (dashed) 
at fixed Fermi energies
are plotted as a function of $1/W$ (not $W$).
In the weak disorder regime ($W<W_{\rm c}$), the SCBA conductivity 
is proportional to $1/W$, and it coincides nicely
with the Boltzmann conductivity Eq.\ (\ref{Boltzmann})
except for a constant shift.
In increasing the disorder (i.e., decreasing $1/W$),
on the other hand, the SCBA conductivity reaches a minimum
at a certain point, and it turns to increase nearly in proportional to $W$. 
The scattering strength for the turning point
is of the order of $W_{\rm c}$, and moves toward larger $W$ 
(i.e., smaller $1/W$) for larger Fermi energy.

Fig.\ \ref{imp}(b) presents a similar plot at
zero energy, where the approximate SCBA solution, 
Eqs.\ (\ref{cond_approx_zero_1}) and (\ref{cond_approx_zero_2}),
is plotted as a dotted-dashed line.
We also show the numerical SCBA conductivity
at $\e =0.001\e_0$, slightly away from the Weyl point.
In the weak disorder regime ($W<W_{\rm c}$),
the SCBA conductivity is very sensitive to $\vare$
as expected from the sharp dip structure in Fig.\ \ref{fe}(d).
The energies $\e=0$ and $\e =0.001\e_0$
correspond to the bottom of the dip and 
the flat region outside the dip, respectively.
The conductivity in $W<W_{\rm c}$ 
is exponentially small at $\e=0$ as already argued,
while at $\e =0.001\e_0$ it linearly rises 
approximately in accordance with 
the analytical expression Eq.\ (\ref{cond_approx_zero_2}).
In the strong disorder regime ($W>W_{\rm c}$),
the SCBA conductivity 
is almost identical
between the two different energies, and goes up
nearly in accordance with 
Eq.\ (\ref{cond_approx_zero_1}).

In 3D Weyl electron, 
the Weyl-point conductivity is highly $W$-dependent
since Eq.\ (\ref{cond_app_e0}) is proportional to $\Gamma(0)$, 
and it abruptly rises when $W$ exceeds $W_{\rm c}$
just in the same way as the density of states.
This is in  a sharp contrast to the 2D case,
where the Weyl-point conductivity becomes nearly 
universal value of the order of $e^2/h$.
\cite{ludwig1994integer,shon1998quantum,ziegler1998delocalization,katsnelson2006zitterbewegung,tworzydlo2006sub,noro2010theory,fradkin1986critical2}
The 3D Weyl system does not have
such a universal conductivity,
because the conductivity in 3D has a dimension of 
$e^2/h$ times the inverse of the length scale,
and this is given by $\Gamma(0)/(\hbar v)$ in the present system.



\section{Conclusion}
\label{sec_conc}

We have studied the electronic transport in disordered 
three-dimensional Weyl electron system 
using the self-consistent Born approximation.
The scattering strength is characterized by
the dimensionless parameter $W$ determined by the scattering amplitude,
and the conductivity significantly changes its behavior
at the certain scattering strength $W_{\rm c}\simeq 1.806$.
In the weak disorder regime $(W<W_{\rm c})$, 
the SCBA conductivity off the Weyl point 
mostly agrees with the Boltzmann conductivity
which is proportional to $1/W$, while 
in the strong disorder regime  $(W>W_{\rm c})$,
the conductivity becomes larger in increasing the disorder
contrary to the usual metallic behavior.
The conductivity at Weyl point is not universal unlike in 2D,
but highly $W$-dependent just in the same way as the density of states.
It is exponentially small in the weak disorder regime,
and abruptly rises when $W$ exceeds $W_{\rm c}$.

Throughout the paper, we assumed a single-node 3D Weyl Hamiltonian
with Gaussian impurity scatterers.
We expect that the theory applies to the Weyl semimetals as long as
different Weyl nodes are well separated in $k$-space,
and the disorder potential is sufficiently smooth 
not to mix up the different nodes.
The valley mixing effect should be important when
two or more Weyl nodes are degenerate in $k$-space.
In the 2D Weyl electron system, it was reported that 
the valley mixing effect does not change the qualitative behavior
of the SCBA conductivity \cite{shon1998quantum},
while the same problem in 3D requires 
a further investigation.

The dependence of the transport property
on the specific form of impurity potential $u(q)$ 
is also an important problem.
In Gaussian impurities, the existence of
the critical disorder strength $W_{\rm c}$ is attributed to the fact that
the self-consistent equation Eq.\ (\ref{eq_self_app_e0})
only allows the trivial solution $\Gamma=0$ in $W < W_{\rm c}$,
and this restriction is imposed by
the upper limit of the integral in Eq.\ (\ref{eq_self_app_e0_2}).
The existence of $W_{\rm c}$ for other long range impurities
should be examined by an detailed SCBA calculation,
while it is roughly estimated by a similar argument.
For Coulomb impurities $u(q) \propto 1/q^2$, for example,
the squared matrix element $u(q)^2$ 
gives an extra $1/k^4$ term in $k$-integral in Eq.\ (\ref{eq_self_app_e0_2}).
Then there is no upper limit in the integral, so that 
we always have non-zero solution for $\Gamma$, giving no $W_{\rm c}$.
We expect that $W_{\rm c}$ exists in the type of impurity
such that $u(q)$ remains finite at $q=0$.

A 3D Weyl electron system in condensed matter
always has the end of the linear dispersion in high energy, 
and this may affect the transport property.
In this work, we showed that 
the SCBA conductivity in Gaussian impurities is 
well described by an analytic approximation,
where $u(q)$ is approximated by a constant and a 
$k$-space cut-off $k_{\rm c}$.
On the contrary, if we regard the end of the linear band
in a real system as a cut-off, 
it would be effectively 
described by the present theory with an appropriate $k_{\rm c}$.

{
{\it Note added}. We recently became aware of recent works
\cite{PhysRevLett.112.016402,biswas2013diffusive}
which predict the rise the density of states 
at the critical disorder strength in the 3D Weyl electron system.
}

\section*{ACKNOWLEDGMENTS}

This project has been
funded by JSPS Grant-in-Aid for Scientific Research No.
24740193 and No. 25107005.

\bibliography{3d_Dirac_dc_cond}

\end{document}